\documentclass[a4paper]{jpconf}
\usepackage{graphicx}
\usepackage{array}
\usepackage{rotating}
\usepackage{multirow}
\makeatletter

\providecommand{\tabularnewline}{\\}

\usepackage[table]{xcolor}
\makeatother
\usepackage{import}

\usepackage{tikz}
\usetikzlibrary{quotes,arrows.meta}
\usetikzlibrary{positioning}

\def\edgecolor{rgb:blue,4;red,1;green,4;black,3}

\usepackage{Ball}
\usepackage{Box}
\usepackage{RightBandedBox}

\usepackage{tikz}
\usetikzlibrary{positioning}
\usetikzlibrary{3d}
\usetikzlibrary{calc}
\usepackage{subcaption}
\usepackage{amsmath}
\usepackage{nicefrac}
\usepackage{adjustbox}

\makeatletter
\def\endthebibliography{%
  \def\@noitemerr{\@latex@warning{Empty `thebibliography' environment}}%
  \endlist
}
\makeatother
\definecolor{grey20}{RGB}{225, 225, 225}
\definecolor{grey40}{RGB}{179, 179, 179}
\definecolor{grey60}{RGB}{97, 98, 101}
\definecolor{gold}{RGB}{244, 170, 0}
\definecolor{copper}{RGB}{215, 118, 0}
\definecolor{652C}{RGB}{114,153,198}
\definecolor{1675C}{RGB}{185,89,21}
\definecolor{teal}{RGB}{0, 129, 171}
\definecolor{green}{HTML}{007836}
\definecolor{yellow}{RGB}{215, 170, 0}
\definecolor{red}{HTML}{BE0F34}
\definecolor{gold}{HTML}{F4AA00}
\definecolor{platinum}{HTML}{B3B3B3}

\newcommand{\hvd}{\texttt{Horovod}}
\newcommand{\torch}{\texttt{Torch}}
\newcommand{\matex}{\texttt{MaTEx}}
\newcommand{\summitdev}{\textit{SummitDev}}

\begin{document}
\title{Scaling the training of particle classification on simulated MicroBooNE events to multiple GPUs}

\author{A Hagen, E Church, J Strube, K Bhattacharya, and V Amatya}
\address{Pacific Northwest National Laboratory, Richland, WA, USA}
\ead{alexander.hagen@pnnl.gov}

\begin{abstract}
  Measurements in Liquid Argon Time Projection Chamber (LArTPC) neutrino detectors, such as the MicroBooNE detector at Fermilab \cite{Acciarri2017}, feature large, high fidelity event images. Deep learning techniques have been extremely successful in classification tasks of photographs, but their application to LArTPC event images is challenging, due to the large size of the events. Events in these detectors are typically two orders of magnitude larger than images found in classical challenges, like recognition of handwritten digits contained in the MNIST database or object recognition in the ImageNet database. Ideally, training would occur on many instances of the entire event data, instead of many instances of cropped regions of interest from the event data. However, such efforts lead to extremely long training cycles, which slow down the exploration of new network architectures and hyperparameter scans to improve the classification performance.
  We present studies of scaling a LArTPC classification problem on multiple architectures, spanning multiple nodes. The studies are carried out on simulated events in the MicroBooNE detector.  We emphasize that it is beyond the scope of this study to optimize networks or extract the physics from any results here. Institutional computing at Pacific Northwest National Laboratory and the \textit{SummitDev} machine at Oak Ridge National Laboratory’s Leadership Computing Facility have been used. To our knowledge, this is the first use of state-of-the-art Convolutional Neural Networks for particle physics and their attendant compute techniques onto the DOE Leadership Class Facilities. We expect benefits to accrue particularly to the Deep Underground Neutrino Experiment (DUNE) LArTPC program, the flagship US High Energy Physics (HEP) program for the coming decades.
\end{abstract}

\section{Introduction\label{sec:introduction}}

Use of convolutional networks to analyze time projection chamber data is often performed on cropped data because of large image sizes.  Training and inference on uncropped TPC data is desired to minimize physics information loss before training.  The high fidelity and large size of the image data requires scaling of computing resources past the 1s to 10s and 100s of GPUs.

\subsection{The MicroBooNE Detector and data format used}

This work formats its simulated data with inspiration from the MicroBooNE experiment \cite{Acciarri2017}.  MicroBooNE is a $170\;\mathrm{tonne}$ liquid argon time projection chamber (LArTPC) with the express interest of analyzing neutrino physics.  Readout of MicroBooNE consists of 2 induction planes with $2400\;\mathrm{wires}$ each and 1 collection plane with $3156\;\mathrm{wires}$.  Readout occurs every $4.8\;\mathrm{ms}$ (which is $2.2\times$ the TPC drift time) for $9600\;\mathrm{digitizations}$.

\begin{figure}
  \centering
  \adjustbox{width=0.98\textwidth}{
    \input{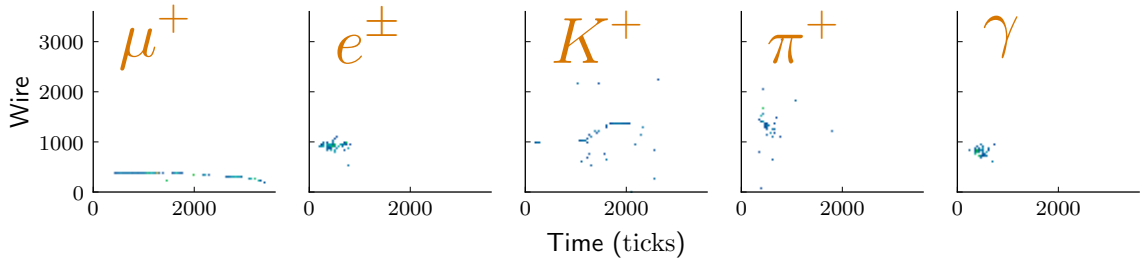}
  }

  \caption{Example of simulated single particle interactions for classification.}\label{fig:events}
\end{figure}

This work presents classification on simulated single particle events in a format similar to that of MicroBooNE \footnote{Permission from MicroBooNE collaboration was granted to focus on compute resources and performance studies}.  GEANT4 Monte Carlo particle transport code was used to simulate interactions with the wires in a LArTPC.  The seed particles were single particles of type $\mu^{+}$, $e^{+}$, $e^{-}$, $K^{+}$, $\pi^{+}$, and $\gamma$.  The interactions with the collection plane were then tallied across $3200$ time ticks and padded with zeros to increase the time dimension to $3600$.  The collection plane dimension was also padded with zeros to reach size $3600$.  The final simulated data was a $3600\times 3600$ "image", with examples for each particle shown in Figure \ref{fig:events}.

Previous work showed not only decreased training time for simple convolution networks on 2 to 14 GPUs (on PNNL's Institutional Computing), but improved performance, achieving lower losses with 14 GPUs versus with only 1 \cite{Bhattacharya2018}.  This work extends upon that scaling study, moving from 1 to 10s of GPUs to the 100s.

\section{Methodology}

\begin{figure}
  \centering
  \adjustbox{width=0.50\columnwidth}{
    \input{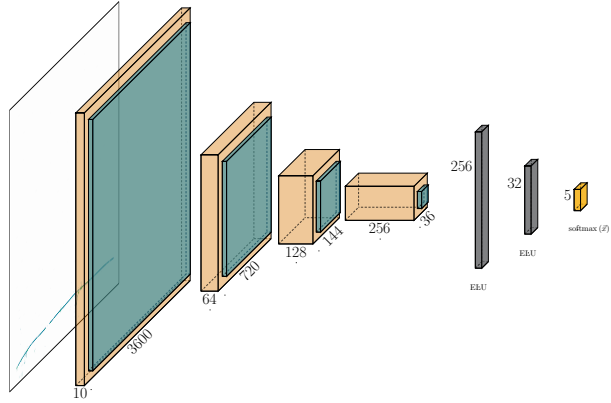}
  }

  \caption{Design of "\texttt{JishNet}" and "\texttt{SCNet}" convolutional networks used for all testing presented on this poster\label{fig:network-architecture}}
\end{figure}

A simplified convolutional neural network (CNN) model was developed for testing\footnote{It is stressed that this model was generated to provide model results while focusing on studying the computatonal resources.  The CNN itself could be improved in numerous ways, but that is past the scope of this work.}. As shown in figure \ref{fig:network-architecture}, the network architecture used two subsequent blocks consisting of convolutional layers with kernel size 5, padding of 2, and exponential linear unit (ELU) activation; followed by max pool layers of pooling size 5. Then, two more subsequent blocks of convolutional layers with kernel size 4, padding of two, and ELU activation followed by max pooling layers of pool size 4.  These four blocks have increasing numbers of convolutional filters with the scheme of 1, 10, 64, 128, 256.  After these four blocks, a linear layer with 20736 inputs, 32 outputs and ELU activation was appended. A linear layer with 32 inputs, 5 outputs, and softmax activation, one output for each of the particle types ($e^{+}$ and $e^{-}$ were grouped together), finished the network.  For a dense data representation, this network architecture was implemented using \texttt{Torch}'s \texttt{Conv2d} and \texttt{MaxPool2d} layers and is hereafter called \texttt{JishNet}. For a sparse data representation, discussed further in section \ref{sec:sparse}, this network used \texttt{SparseConvNet}'s \texttt{SubmanifoldConvolution} and \texttt{MaxPooling} layers and is hereafter referred to as \texttt{SCNet}. It should be noted that the padding behavior of the convolutions in \texttt{JishNet} and \texttt{SCNet} differ subtly.

\subsection{Resources}

Many industrial applications of convolutional networks can apply 10s to 100s of GPUs to speed training and enable fast iteration in network design \cite{Mikami2018, Sergeev2018}.

\subsubsection{Hardware}

This work presents one of the first known instances of scaling a physics classificaton problem to this scale on Oak Ridge National Laboratory Leadership Computing Facility's \summitdev{} computer. \summitdev{}, which is a test stand for the eventual \textit{Summit} computer, was used in this study.  The \summitdev{} machine boasts 50 nodes, each with 2 22-core IBM Power8NVL CPUs and 4 NVIDIA P100 GPUs.  The interconnections between nodes are Mellanox EDR 100G InifiBand; interconnections between GPU to host node are NVLink.  It should also be noted that there is a 4 hour time limit to all \summitdev{} jobs.

\subsubsection{Software}

There is an embarassment of options for the development of neural networks and data parallel training of these neural networks.  While previous scaling studies used Pacific Northwest National Laboratory's \matex{}  \cite{Bhattacharya2018,Amatya2018} code for data parallel training, this work was based on the use of \torch{} \cite{Paszke2017} for neural network definition and training, and \hvd{} \cite{Sergeev2018} for data parallelization.

Due to the large size of images and their large number, compression was needed to store the dense representation of the images in memory when loading\footnote{No compression was needed to store the sparse representation of these events.}.  Compression using \texttt{Blosc} \cite{Alted2019} when loading images from file into CPU memory.  At training time, a minibatch of these images were uncompressed from memory and transferred to GPU memory.  Due to the large size of each datum, only 7 images alongsize the \texttt{JishNet} model could fit into a single NVIDIA P100's memory.  Thus, all training was performed with a minibatch size of $7$.  After training of a single minibatch on all GPUs allocated for each training instance, \hvd{}'s \texttt{allreduce} function was used to transfer and allocate all gradients to the head node.  This has the effect of generating a batch size of $7 \times N_{GPU}$.  The loss and accuracy local to each GPU was written to disk before this \texttt{allreduce} operation.

The use of many GPUs significantly affects the training dynamics of the CNN in this training example.  A prolonged study was performed examining the training dynamics of this task using stochastic gradient descent (SGD) optimizer and warmup and scaling suggested in Goyal \cite{Goyal2017}.  The SGD optimization study ultimately underperformed, so other options were explored. Distributed Adam optimization was tested and ultimately successful; thereafter, all results used Adam optimization \cite{Kingma2014}. The training dynamics were highly sensitive to parameters of Adam optimizer, final parameters were $lr=0.001$, $\beta = \left[0.9, 0.999\right]$, $\epsilon = 1\times 10^{-8}$, and $decay = 0.01$.  Figures \ref{fig:jishnet_loss} and \ref{fig:scnet_loss} illustrate expected training behavior and verify the use of the Adam optimizer.

\subsection{Sparse\label{sec:sparse}}

Most of the data passed to the dense convolutional network is zeros, as no events happen in large parts of the TPC. Sparse convolutional networks are a more efficient network architecture for LArTPC event classificaton. After conversion from a dense format to a sparse data format, training using \texttt{SubmanifoldConvolution} layers from Facebook's \texttt{SparseConvNet} \cite{Graham2017} codebase and the modified CNN \texttt{SCNet} was performed.

\section{Results}

Results show increasing the number of GPUs for both dense (\texttt{JishNet}) and sparse (\texttt{SCNet}) CNNs was successful in decreasing the training time on large images in large datasets.

\subsection{Dense}

\begin{figure}
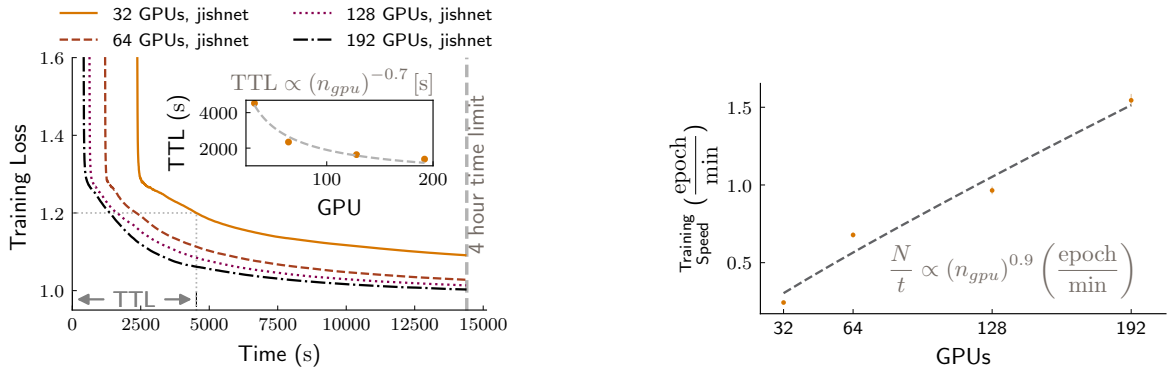

  \centering
  \begin{subfigure}{0.45\textwidth}
    \adjustbox{width=1.0\textwidth}{
      \input{img/train_loss_vs_time_jishnet.pgf}
    }
    \caption{Training dynamics of \texttt{JishNet} using increasing number of GPUs on \summitdev{}.}\label{fig:jishnet_loss}
  \end{subfigure}
  \hfill%
  \begin{subfigure}{0.45\textwidth}
    \vspace{3em}%
    \adjustbox{width=1.0\textwidth}{
      \input{img/train_time_vs_epochs.pgf}
    }
    \caption{Speedup of training \texttt{JishNet} with increasing number of GPUs on \summitdev{}.}\label{fig:jishnet_speedup}
  \end{subfigure}
  \vspace{1em}%

  \caption{Results from training the dense convolutional network \texttt{JishNet} using various numbers of GPUs on \summitdev{}.}%
  \label{fig:training_jishnet}
\end{figure}

Scaling to multiple GPUs shows an approximately linear speedup with respect to the number of GPUs, as shown in Figure \ref{fig:jishnet_speedup}.  Figure \ref{fig:jishnet_loss} shows the loss versus training wall time with increasing GPUs.  With more GPUs, a lower loss can be achieved in the allotted 4 hours, and the time to a loss (TTL) of $1.2$ decreases with the number of GPUs used (shown in inset).

In Figure \ref{fig:jishnet_loss}, there is a marked difference in the time at which the first loss was reported with different number of GPUs.  This time, the data loading time (DLT), is affected by how many images must be loaded into CPU before beginning training.  Each training example was performed on 15,000 images, so each CPU had to load and compress $\nicefrac{15,000}{N_{GPU}}$ into memory.  Optimization of this compression routine is left for future work.  It is hypothesized that the TTL speedup is sublinear ($\propto N_{GPU}^{0.7)}$) because of this DLT effect.

\subsection{Sparse}

\begin{figure}
  \centering
  \begin{subfigure}{0.45\textwidth}
    \adjustbox{width=1.0\textwidth}{
      \input{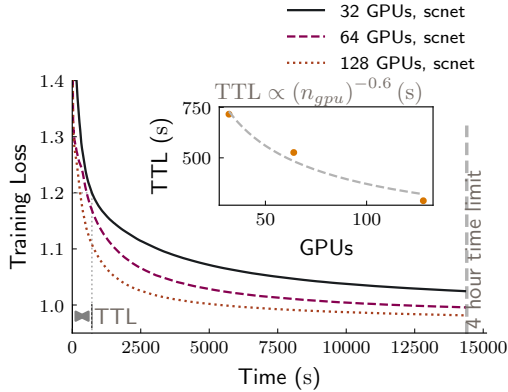}
    }
    \caption{Training dynamics of \texttt{SCNet} using increasing number of GPUs on \summitdev{}.}\label{fig:scnet_loss}
  \end{subfigure}
  \hfill%
  \begin{subfigure}{0.45\textwidth}
    \vspace{3em}%
    \adjustbox{width=1.0\textwidth}{
      \input{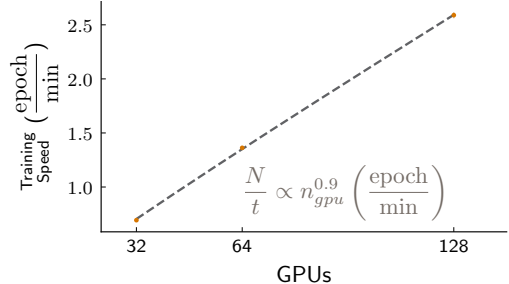}
    }
  \caption{Speedup of training \texttt{SCNet} with increasing number of GPUs on \summitdev{}.}\label{fig:scnet_speedup}
  \end{subfigure}

  \vspace{1em}%

  \caption{Results from training the sparse convolutional network \texttt{SCNet} using various numbers of GPUs on \summitdev{}.}%
  \label{fig:training_scnet}
\end{figure}

\texttt{SCNet} shows interesting behavior during training compared to \texttt{JishNet}.  Figure \ref{fig:scnet_speedup} again shows an approximately linear speedup when trained on increasing numbers of GPUs, however the highest speed is $\sim 2.5\;\nicefrac{\mathrm{epochs}}{\mathrm{min}}$, which is $60\;\mathrm{\%}$ faster than \texttt{SCNet}.  However, the TTL, shown in Figures \ref{fig:scnet_loss} is $4\times$ lower than that in Figure \ref{fig:jishnet_loss}.

Figure \ref{fig:scnet_loss} shows comparable loss to \texttt{JishNet} after the full training set, however there are several interesting characteristics.  Compared to Figure \ref{fig:jishnet_loss}, there is very little evident DLT in Figure \ref{fig:scnet_loss}.  Despite this, the TTL speedup is only $\propto N_{GPU}^{0.6}$.

It should be noted that the 7 image GPU memory limitation did not apply for \texttt{SCNet}, due to the small size of each datum, yet we kept the minibatch size at 7, nevertheless.  This study did not explore varying the minibatch size (and in extension batch size), but the researchers hypothesize that increased speedup could be realized using more optimal batch sizes during training of \texttt{SCNet}.

\section{Discussion and Conclusions}

\begin{table}
  \centering
  \caption{Confusion matrices from training of \texttt{JishNet} and \texttt{SCNet} with 128 GPUs for 4 hours.}%
  \label{tab:cm}%
  \vspace{1em}%
  \begin{subtable}{0.45\textwidth}
  \caption{Confusion matrix for \texttt{JishNet} trained with 15000 images for 4 hours on 128 GPUs.}\label{tab:cm_dense}
    \hspace{-0.325\textwidth}
    \adjustbox{width=1.25\textwidth}{
      
\definecolor{color11}{HTML}{BBDBCA}
\definecolor{color12}{HTML}{F6DFE4}
\definecolor{color13}{HTML}{FFFFFF}
\definecolor{color14}{HTML}{FFFFFF}
\definecolor{color15}{HTML}{FFFFFF}
\definecolor{color21}{HTML}{FFFDFE}
\definecolor{color22}{HTML}{9BCAB1}
\definecolor{color23}{HTML}{FFFFFF}
\definecolor{color24}{HTML}{FFFFFF}
\definecolor{color25}{HTML}{FFFFFF}
\definecolor{color31}{HTML}{FFFFFF}
\definecolor{color32}{HTML}{FFFFFF}
\definecolor{color33}{HTML}{9ACAB0}
\definecolor{color34}{HTML}{FFFFFF}
\definecolor{color35}{HTML}{FFFEFE}
\definecolor{color41}{HTML}{FFFDFE}
\definecolor{color42}{HTML}{FFFDFE}
\definecolor{color43}{HTML}{FFFEFE}
\definecolor{color44}{HTML}{D9EBE1}
\definecolor{color45}{HTML}{F0C7D0}
\definecolor{color51}{HTML}{FFFEFF}
\definecolor{color52}{HTML}{FFFFFF}
\definecolor{color53}{HTML}{FEFDFD}
\definecolor{color54}{HTML}{F8E6EA}
\definecolor{color55}{HTML}{B7D9C6}
\begin{tabular}{cc|ccccc}
 &  & \multicolumn{5}{c}{\textbf{Predicted}}\tabularnewline
 &  & $\gamma$ & $e \pm$ & $\mu +$ & $\pi +$ & $K +$\tabularnewline
\hline
\multirow{5}{*}{
\rotatebox{90}{\textbf{True}}
} & $\gamma$ & \cellcolor{color11}$66.7\%$ %
& \cellcolor{color12}$33.0\%$ %
& \cellcolor{color13}$0.0\%$ %
& \cellcolor{color14}$0.0\%$ %
& \cellcolor{color15}$0.3\%$ \tabularnewline
 & $e \pm$ & \cellcolor{color21}$1.6\%$ %
& \cellcolor{color22}$98.0\%$ %
& \cellcolor{color23}$0.0\%$ %
& \cellcolor{color24}$0.4\%$ %
& \cellcolor{color25}$0.0\%$ \tabularnewline
 & $\mu +$ & \cellcolor{color31}$0.0\%$ %
& \cellcolor{color32}$0.1\%$ %
& \cellcolor{color33}$98.9\%$ %
& \cellcolor{color34}$0.0\%$ %
& \cellcolor{color35}$1.0\%$ \tabularnewline
 & $\pi +$ & \cellcolor{color41}$1.6\%$ %
& \cellcolor{color42}$1.7\%$ %
& \cellcolor{color43}$1.3\%$ %
& \cellcolor{color44}$37.5\%$ %
& \cellcolor{color45}$57.8\%$ \tabularnewline
 & $K +$ & \cellcolor{color51}$0.6\%$ %
& \cellcolor{color52}$0.1\%$ %
& \cellcolor{color53}$2.1\%$ %
& \cellcolor{color54}$26.3\%$ %
& \cellcolor{color55}$71.0\%$ \tabularnewline
\end{tabular}

    }
  \end{subtable}
  \hfill%
  \begin{subtable}{0.45\textwidth}
  \caption{Confusion matrix for \texttt{SCNet} trained with 15000 images for 4 hours on 128 GPUs.}\label{tab:cm_sparse}
    \hspace{-0.325\textwidth}
    \adjustbox{width=1.25\textwidth}{
      
\definecolor{color11}{HTML}{BCDBCA}
\definecolor{color12}{HTML}{F6DFE4}
\definecolor{color13}{HTML}{FFFFFF}
\definecolor{color14}{HTML}{FFFEFE}
\definecolor{color15}{HTML}{FFFFFF}
\definecolor{color21}{HTML}{FFFEFE}
\definecolor{color22}{HTML}{9ACAB0}
\definecolor{color23}{HTML}{FFFFFF}
\definecolor{color24}{HTML}{FFFFFF}
\definecolor{color25}{HTML}{FFFFFF}
\definecolor{color31}{HTML}{FFFFFF}
\definecolor{color32}{HTML}{FFFFFF}
\definecolor{color33}{HTML}{9AC9AF}
\definecolor{color34}{HTML}{FFFFFF}
\definecolor{color35}{HTML}{FFFFFF}
\definecolor{color41}{HTML}{FFFFFF}
\definecolor{color42}{HTML}{FFFEFE}
\definecolor{color43}{HTML}{FFFEFE}
\definecolor{color44}{HTML}{BADAC9}
\definecolor{color45}{HTML}{F7E2E7}
\definecolor{color51}{HTML}{FFFFFF}
\definecolor{color52}{HTML}{FFFFFF}
\definecolor{color53}{HTML}{FFFDFE}
\definecolor{color54}{HTML}{EEBFC9}
\definecolor{color55}{HTML}{E0EFE7}
\begin{tabular}{cc|ccccc}
 &  & \multicolumn{5}{c}{\textbf{Predicted}}\tabularnewline
 &  & $\gamma$ & $e \pm$ & $\mu +$ & $\pi +$ & $K +$\tabularnewline
\hline
\multirow{5}{*}{
\rotatebox{90}{\textbf{True}}
} & $\gamma$ & \cellcolor{color11}$66.1\%$ %
& \cellcolor{color12}$33.2\%$ %
& \cellcolor{color13}$0.0\%$ %
& \cellcolor{color14}$0.7\%$ %
& \cellcolor{color15}$0.0\%$ \tabularnewline
 & $e \pm$ & \cellcolor{color21}$0.9\%$ %
& \cellcolor{color22}$98.8\%$ %
& \cellcolor{color23}$0.0\%$ %
& \cellcolor{color24}$0.3\%$ %
& \cellcolor{color25}$0.0\%$ \tabularnewline
 & $\mu +$ & \cellcolor{color31}$0.0\%$ %
& \cellcolor{color32}$0.0\%$ %
& \cellcolor{color33}$99.4\%$ %
& \cellcolor{color34}$0.0\%$ %
& \cellcolor{color35}$0.5\%$ \tabularnewline
 & $\pi +$ & \cellcolor{color41}$0.0\%$ %
& \cellcolor{color42}$1.1\%$ %
& \cellcolor{color43}$1.0\%$ %
& \cellcolor{color44}$68.0\%$ %
& \cellcolor{color45}$29.9\%$ \tabularnewline
 & $K +$ & \cellcolor{color51}$0.4\%$ %
& \cellcolor{color52}$0.4\%$ %
& \cellcolor{color53}$1.8\%$ %
& \cellcolor{color54}$67.1\%$ %
& \cellcolor{color55}$30.4\%$ \tabularnewline
\end{tabular}

    }
  \end{subtable}
\end{table}

Training both dense and sparse convolutional networks on 10s to 100s of GPUs proved successful in this study.  The wall time to a training loss decreased linearly with increased amounts of GPU resources, and allowed training on large images for physics related classificatoin studies.  The final accuracy metrics for both of these networks are shown in Tables \ref{tab:cm_dense} and \ref{tab:cm_sparse}. The final validation accuracy for \texttt{JishNet} was $78.8\% \pm 3.8\%$, and for \texttt{SCNet} was $76.8\% \pm 3.3\%$. Particle labeling confusion is observed among species for which physics intuition suggests it is not unexpected.  This study also elucidated several other important aspects related to training CNNs with large image sizes.  These aspects included the sensitivity of training to the optimizer parameters and seeming inappropriateness of suggestions in the literature for training with SGD. We also uncovered the importance of minibatch size and tradeoff between GPU memory resources and optimal training dynamics.

\section*{Acknowledgements}

The authors gratefully acknowledge the MicroBooNE collaboration for permission to work on simulated LArTPC data to focus on compute resources and performance scaling. This research used resources of the Oak Ridge Leadership Computing Facility at the Oak Ridge National Laboratory, which is supported by the Office of Science of the U.S. Department of Energy under Contract No. DE-AC05-00OR22725. Thus, the Oak Ridge Leadership Computing Facility and its staff are also gratefully acknowledged for use of the \summitdev{} computer.  The authors also gratefully acknowledge Pacific Northwest National Laboratory's Institutional Computing for use of its resources.

\section*{References}
\bibliography{acat_2019}

\end{document}